\documentclass[%
 aip,
 amsmath,amssymb,
 reprint,%
]{revtex4-1}
\usepackage{graphicx}  
\usepackage{dcolumn}   
\usepackage{bm}        
\usepackage{amssymb}   
\usepackage[utf8]{inputenc}
\usepackage[T1]{fontenc}
\usepackage{mathptmx}
\usepackage{mathtools}
\usepackage{hyperref}
\usepackage{cleveref}
\hypersetup{colorlinks=true, citecolor=blue, urlcolor=blue, linkcolor=blue}

\begin{document}



\title{Non-Adiabatic Effects on Electron Beam Quality for Frequency-Tunable Gyrotrons}
%
%
\author{Cheng-Hung~Tsai} 
 \affiliation{Department of Physics, National Tsing Hua University, Hsinchu 30013, Taiwan}
\author{Toshitaka~Idehara}
\author{Yuusuke~Yamaguchi} 
 \affiliation{Research Center for Development of Far-Infrared Region, University of Fukui (FIR UF), Fukui 910-8507, Japan}
\author{Tsun-Hsu~Chang}
 \email{thschang@phys.nthu.edu.tw}
 \affiliation{Department of Physics, National Tsing Hua University, Hsinchu 30013, Taiwan}

\date{\today}

\begin{abstract}
We propose an unconventional electron gun structure in which the emitter is located on a concave cathode surface with a non-uniform electric field.
Such a design violates the intuition that an emitter should place close to a uniform electric field to reduce the velocity spread.
The commonly employed design guide based on the adiabatic condition predicts a huge velocity spread of 24$\%$, but the simulation using EGUN code and verified with CST particle studio shows a very low spread of 2.8$\%$.
Examining the magnetic moment and the kinetic energy of the beam reveals that the electrons experience a relatively long acceleration process due to the much weak electric field.
That's why the non-adiabatic effect matters.
In addition to the cyclotron compression and the $\textbf{E}\times\textbf{B}$ drift, the ``resonant'' polarization drift plays a crucial role in reducing the overall velocity spread.
Simulations show a decent beam quality with the pitch factor of 1.5 and the transverse velocity spread of 2.8$\%$ over a wide range of the magnetic field ($7.4-8.0$ T) and the beam voltage ($12-22$ kV) with a high structural tolerance.
The promising results with the wide working range enable the development of continuous frequency-tunable gyrotrons.
\end{abstract}

\maketitle


Gyrotrons based on the mechanism of the electron cyclotron maser (ECM) \cite{Chu2004,Nusinovich2004,Thumm2018} is capable of producing high-power, coherent, terahertz (THz) waves as compared with the classical vacuum electronic tubes.
Through decades of development, frequency-tunable gyrotrons receive more and more attentions \cite{Idehara1999,Idehara2007,Miyazaki2014} because of the frequency-sensitive applications, such as the enhancement of nuclear magnetic resonance by dynamic nuclear polarization (DNP-NMR) and the measurement of the hyperfine splitting of positronium (Ps-HFS).
The beam-wave synchronism for gyrotrons is described as:
\begin{equation}
\omega=k_{z}v_{z}+s\Omega_{c},
\label{eq:one}
\end{equation}
where $\omega$ is the angular frequency, $k_{z}$ is the wavenumber, and $s$ is the cyclotron harmonic number.
$\Omega_{c}$ is the relativistic electron cyclotron frequency which is related to the magnetic field ($B_{0}$) and the relativistic gamma factor ($\gamma_{0}$), i.e., $\Omega_{c}=eB_{0}/m_{0}\gamma_{0}$.
$e$ and $m_{0}$ are the charge and the rest mass of an electron. 
$v_{z}$ is the electron axial velocity which is associated with the beam voltage ($V_{b}$).
Equation~(\ref{eq:one}) predicts that the frequency tunability of gyrotrons \cite{Hornstein2005,Chang2009} is attainable by adjusting either $B_{0}$ or $V_{b}$.
Since the output power of gyrotrons is strongly related to the quality of the electron beam, high-performance electron guns deserve in-depth studies for frequency-tunable gyrotrons.

\begin{figure*}
\includegraphics[scale=0.8]{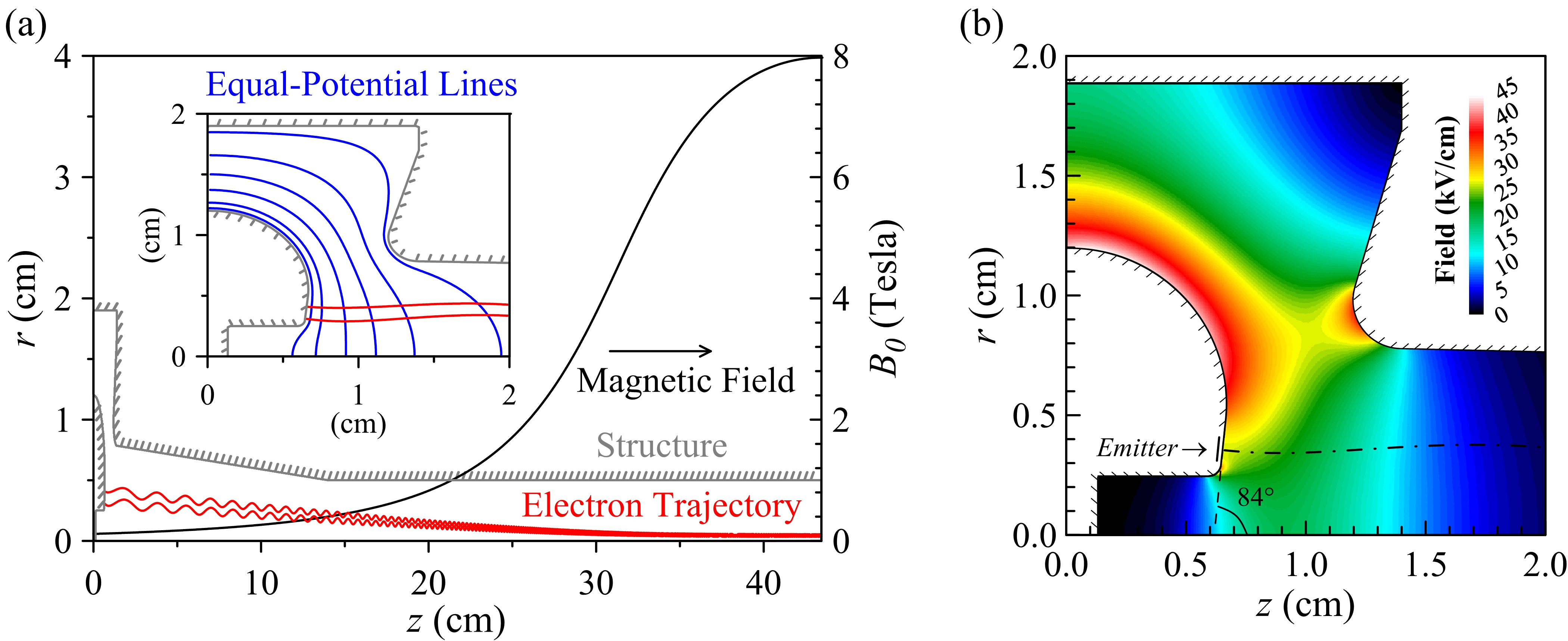}
\caption{\label{fig:Figure1} Sketch of the proposed electron gun.
Structure of the cathode and the anode is drawn in grey lines with slashes.
The magnetic field profile is displayed in a black line and two representative trajectories of electrons are shown in red curves.
The equal-potential lines are depicted in the inset with $V_{b}=16$ kV.
(b) Electric field distribution ($\textit{color scale}$) between electrodes overlaid with the central beam trajectory ($\textit{dotted-dash line}$) from the emitter surface ($\textit{solid line}$).
The emitter is located at a relatively weak electric field, resulting in a much longer acceleration interval.}
\end{figure*}

Many types of electron guns have been proposed \cite{Singh2012,Piosczyk1989,Pikin2016} to generate a high-quality beam with a high pitch factor ($\alpha$) and a low velocity spread ($\Delta v/v$).
Magnetron injection gun (MIG) \cite{Mark1986,Yuan2009,Yamaguchi2012} consisting of a conical-shaped cathode is one of the commonly used electron sources for gyrotrons.
Electrons emitted from a convex surface with a maximal electric field are quickly accelerated to a constant value of energy, i.e., $p^{2}_{\perp}+p^{2}_{z}=(1+\alpha^{2})p^{2}_{z}=const.$.
Then, electrons undergo a magnetic compression according to the adiabatic theory, where $p^{2}_{\perp}/B$ is a constant of motion.
These two invariant quantities imply a relation of the pitch factor:
\begin{equation}
\alpha^2_{2}=\alpha^2_{1}\frac{f}{1+\alpha^2_{1}(1-f)},
\label{eq:two}
\end{equation}
where $f$ is the magnetic compression ratio ($=B_{2}/B_{1}$).
Equation~(\ref{eq:two}) suggests that increasing the magnetic compression ratio enhances the pitch factor of the electron beam under the adiabatic condition.

In this study, we propose a diode-type electron gun to produce a high-quality annular beam.
In contrast to the MIGs, electrons are emitted from a concave cathode surface which is usually used to generate the axis-encircling electron beam (cusp gun) \cite{Du2012,Du2015}. \crefformat{figure}{Figure~#2#1{(a)}#3}\cref{fig:Figure1} shows the proposed structure (\textit{grey}) overlaid with the magnetic field profile (\textit{black}), beam trajectories (\textit{red}), and equal-potential lines (\textit{blue} in the inset) calculated with the beam voltage of 16 kV.
Simulated electric field pattern is displayed in \crefformat{figure}{Fig.~#2#1{(b)}#3}\cref{fig:Figure1} with the maximal field strength of 46 kV/cm.
The emitter is placed, unconventionally, at a weak and non-uniform electric field strength of approximately 25 kV/cm.
The cathode, featuring a nearly planar surface with a large emitter inclination angle ($\theta_{e}$=84$^{\circ}$) and a central notch, is capable of producing a laminar electron beam.
The annular emitter thickness of 1.0 mm suggests a cathode loading ($J_{c}$) of $0.9-4.5$ A/cm$^{2}$ with a beam current of $0.2-1.0$ A.
The maximal cathode loading is smaller than the space-charge limit \cite{Lawson1988} and ensures the lifetime of the emitter.
Simulations are carried out by the particle tracing code (EGUN) and the CST Particle Studio.
Both simulation results agree well.

\begin{figure}
\includegraphics[scale=0.8]{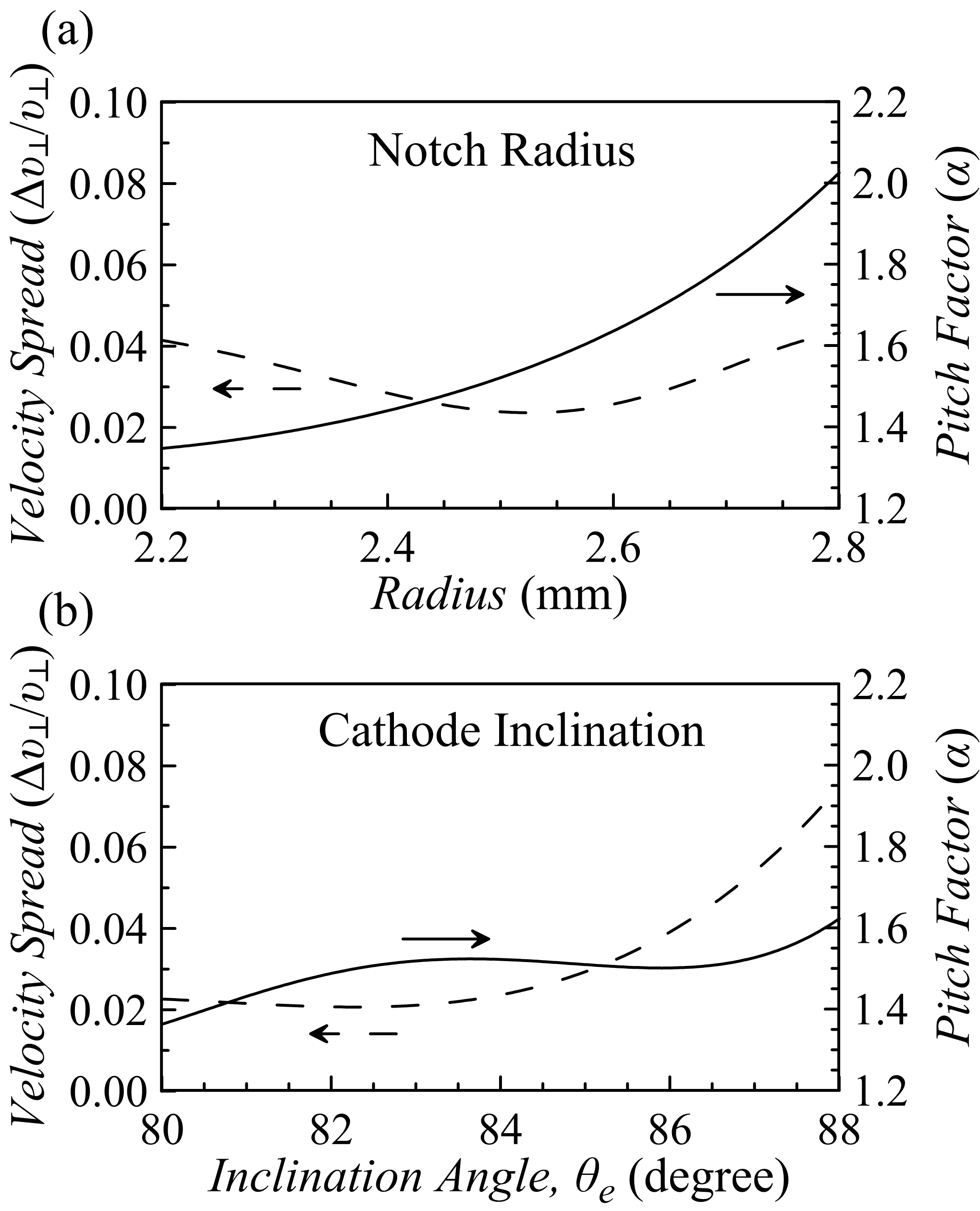}
\caption{\label{fig:Figure2}Sensitivity analysis of two geometric factors: (a) the notch radius and (b) the cathode inclination, on the pitch factor and the transverse velocity spread.}
\end{figure}

Considering the structural tolerance is essential for the design of a high-performance electron gun.
Here we examine the effect of the radius of the central notch and the emitter's inclination angle ($\theta_{e}$).
\crefformat{figure}{Figure~#2#1{(a)}#3}\cref{fig:Figure2} shows the transverse velocity spread and the pitch factor as functions of the notch radius by using the EGUN code.
The larger notch radius results in the higher electric-field strength on the emitter surface due to the edge effect, which also enhance the pitch factor of the electron beam.
The desired pitch factor of 1.5 is obtained at the notch radius of 2.5 mm which also gives the local minimum of transverse velocity spread.
On the other hand, \crefformat{figure}{Fig.~#2#1{(b)}#3}\cref{fig:Figure2} shows the beam quality as a function of the emitter inclination.
The pitch factor changes slightly with different inclination angles while the transverse velocity spread increases gradually with the rising emitter inclination.
The transverse velocity spread shows a low terrace and is insensitive to the inclination angle in a range of 82$^{\circ}$ to 84$^{\circ}$.
In short, \crefformat{figure}{Figure~#2#1{(a)}#3}\cref{fig:Figure2} and \crefformat{figure}{#2#1{(b)}#3}\cref{fig:Figure2} indicate that the beam quality is stable and insensitive to the machining errors.

Reducing the velocity spread of the electron beam is important in improving the output efficiency of gyrotrons.
The axial and transverse velocity spreads are related through the pitch factor under the adiabatic condition \cite{Mark1986}, $\Delta v_{z}/v_{z}=\alpha^{2}(\Delta v_{\perp}/v_{\perp})$.
The velocity spread is traditionally known to be associated with the emitter thermal effect, the emitter surface roughness \cite{Mark1986}, and, most importantly, the fluctuation of the electric and magnetic fields at emitter \cite{Gaponov1981}.
The first two effects are related to the emitter characteristics which contribute a small portion of the velocity spread as compared with the contribution of the inhomogeneous electric and magnetic fields.
Under the adiabatic assumption, an extensively used equation which relates the transverse velocity spread and the fields spread is displayed below \cite{Mark1986}:
\begin{equation}
\frac{\Delta v_{\perp}}{v_{\perp}}=\frac{\Delta E_{\perp}}{E_{\perp}}-\frac{3}{2}\frac{\Delta B_{z}}{B_{z}}.
\label{eq:three}
\end{equation}
However, this relation fails when the non-adiabatic process is non-negligible.
Tables~\ref{tab:table1} shows theoretical predictions based on the adiabatic condition and simulated results.
According to Eq.~(\ref{eq:three}), the large electric field spread ($\Delta E_{\perp}/E_{\perp}\sim24\%$) and negligible magnetic field spread ($\Delta B_{z}/B_{z}<0.01\%$) at emitter suggest a large transverse velocity spread ($\Delta v_{\perp}/v_{\perp}\sim24\%$), which disagrees with simulated results ($\Delta v_{\perp}/v_{\perp}\sim2.8\%$).
Therefore, analyzing the non-adiabatic effect during the electron acceleration may answer the mechanism of why the transverse velocity spread is significantly reduced.

\begin{figure}
\includegraphics[scale=0.8]{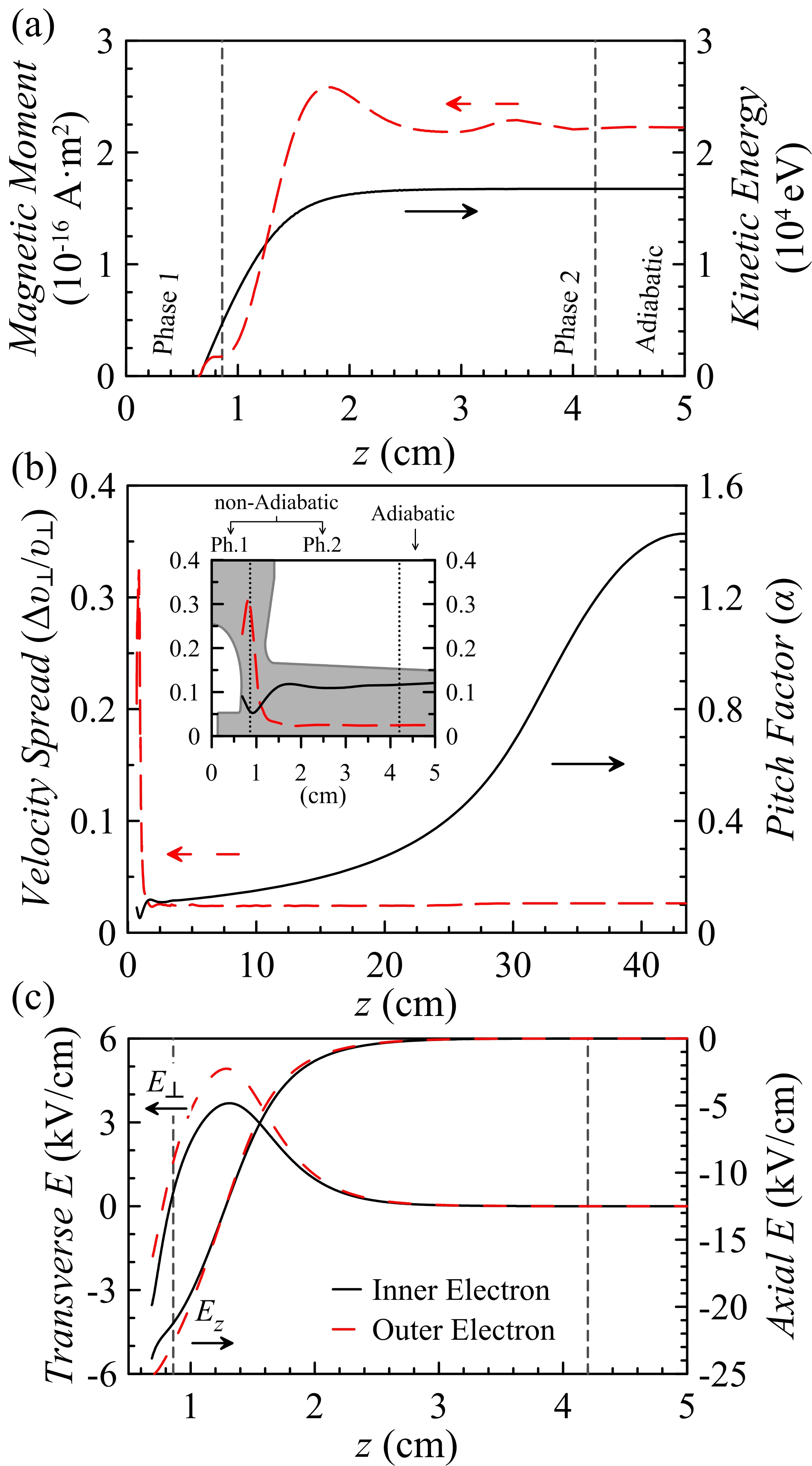}
\caption{\label{fig:Figure3}The key parameters along the axial position based on the EGUN simulation.
(a) The magnetic moment (\textit{dash line}) and the kinetic energy (\textit{solid line}) of the electron beam.
Constants of the magnetic moment and the kinetic energy define the non-adiabatic and the adiabatic zones.
(b) Transverse velocity spread (\textit{dash line}) and the pitch factor (\textit{solid line}) of the electron beam along the z-axis.
The inset enlarges the beam qualities within the space between electrodes (\textit{grey shaded area}).
The surge and the decrease of the velocity spread are classified as \textit{phase} 1 and \textit{phase} 2, respectively.
(c) Transverse and axial electric field strengths experienced by the inner and outer electrons.}
\end{figure}

\crefformat{figure}{Figure~#2#1{(a)}#3}\cref{fig:Figure3} displays the magnetic moment ($\gamma m_{0}v^{2}_{\perp}/2B$, left vertical axis) and the kinetic energy ($(\gamma-1)m_{0}c^{2}$, right vertical axis) as functions of the axial position $z$.
The gradually increased kinetic energy of electrons indicates the acceleration or ``non-adiabatic'' process.
On the other hand, constant kinetic energy implies the ``adiabatic'' condition of electrons without energy exchange.
Changes of the kinetic energy and the magnetic moment of electrons along the z-axis suggest a range of non-adiabatic region before $z= 4.2 \textrm{cm}$.
The magnetic moment of electrons provides more detailed information regarding the non-adiabatic process.
Changes of the magnetic moment in both \textit{phases} are attributed to the cycloidal motion in different directions.
The periodic features in \textit{phase} 2 come from the contribution of non-zero transverse magnetic field \textit{experienced} by electrons in the cyclotron motion.
The magnetic moment gradually reaches a constant value as electrons move to the adiabatic region.

\begin{table}
\caption{\label{tab:table1} The predicted and simulated pitch factors agree well, but the two velocity spreads differ greatly.}
\begin{ruledtabular}
\begin{tabular}{lc}
Magnetic compression ratio, \textit{f} & 67\\
Initial pitch factor, $\alpha_{0}$ & 0.095\\
Predicted pitch factor, $\alpha_{1}$ & 1.5\footnote{based on Eq.~(\ref{eq:two})}\\
Simulated pitch factor, $\alpha$ & 1.5\\
Electric field variation, $\Delta E_{\perp}/E_{\perp}$ & 24$\%$\\
Magnetic field variation, $\Delta B_{z}/B_{z}$ & $<0.01\%$\\
Predicted velocity spread, $\Delta v_{1\perp}/v_{1\perp}$ & $24\%$\footnote{based on Eq.~(\ref{eq:three})}\\
Simulated velocity spread, $\Delta v_{\perp}/v_{\perp}$ & $2.8\%$\\
\end{tabular}
\end{ruledtabular}
\end{table}

\crefformat{figure}{Figure~#2#1{(b)}#3}\cref{fig:Figure3} displays the transverse velocity spread and the pitch factor of the electron beam as functions of the axial position $z$.
The beam quality shows dramatic changes during the non-adiabatic process and becomes stable when electrons reach the adiabatic condition.
Here we only focus on the non-adiabatic behavior of electrons because there are extensive studies on the adiabatic behaviors \cite{Mark1986,Tsimring2007}.
As electrons moving forward, the transverse velocity spread rises from 24$\%$ to 32$\%$ in \textit{phase} 1 of the non-adiabatic region while the pitch factor decreases from 0.095 to 0.05.
However, in \textit{phase} 2, the transverse velocity spread declines rapidly from 32$\%$ to 2.8$\%$ while the pitch factor rises from 0.05 to 0.12 and gradually increases afterward.
\crefformat{figure}{Figure~#2#1{(c)}#3}\cref{fig:Figure3} displays the electric field \textit{experienced} by electrons generated from the inner side (\textit{solid lines}) and the outer side (\textit{dashed lines}) of the emitter.
The transverse field ($E_{\perp}$) reverses from negative to positive in a range of $-3$ kV/cm to 5 kV/cm during the non-adiabatic process while the axial field ($E_{z}$) gradually increases from $-25$ kV/cm.
All the electric field components approach zero when electrons pass through the acceleration region.
Comparison between \crefformat{figure}{Figure~#2#1{(b)}#3}\cref{fig:Figure3} and \crefformat{figure}{#2#1{(c)}#3}\cref{fig:Figure3} suggest that the non-uniform electric field \textit{experienced} by electrons plays an important role to suppress the transverse velocity spread of the electron beam.
The sudden change of the velocity spread can be physically explained by studying an electron motion with the non-uniform electric and magnetic fields.

\begin{figure}
\includegraphics[scale=0.8]{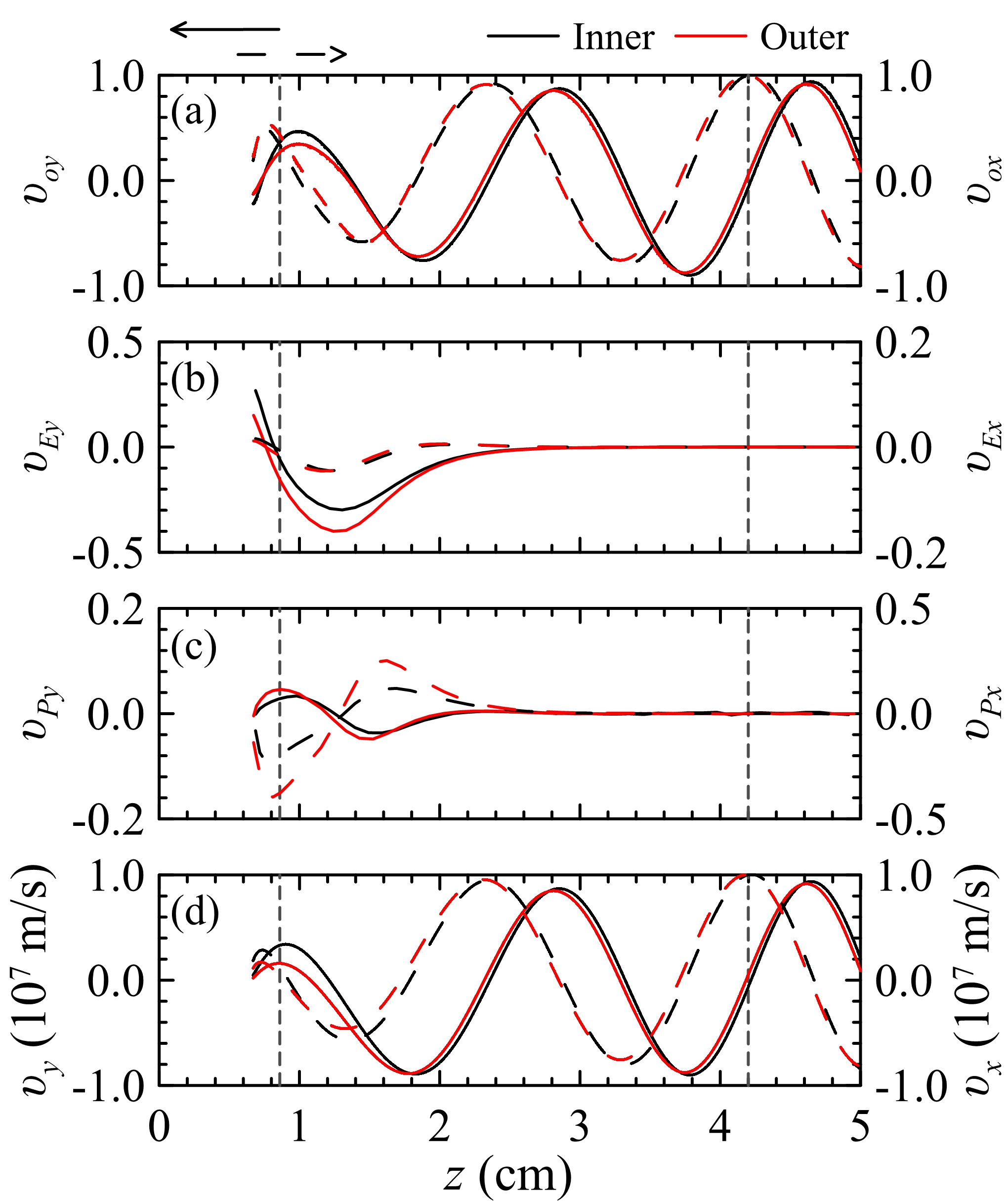}
\caption{\label{fig:Figure4}Transverse velocities of the inner (\textit{black lines}) and the outer (\textit{red lines}) electrons: (a) the cyclotron velocities ($\textbf{v}_{o}$) with the non-uniform magnetic field, (b) the $\textbf{E}\times\textbf{B}$ drift velocities ($\textbf{v}_{E}$), (c) the resonant polarization drift velocities ($\textbf{v}_{p}$), and (d) the resultant velocities based on Eq. (7).}
\end{figure}

By considering the azimuthal symmetry in a cylindrical system, non-uniform electric and magnetic fields can be written as $\textbf{\textrm{E}}=E_{r}\hat{\textbf{r}}+E_{z}\hat{\textbf{z}}$ and $\textbf{\textrm{B}}=B_{r}\hat{\textbf{r}}+B_{z}\hat{\textbf{z}}$.
The non-zero transverse magnetic field is attributed to the fringe effect of the magnet.
The equation of motion of an electron can be described as:
\begin{subequations}
\begin{eqnarray}
&& \dot{v}_{r}=\frac{q}{m}(E_{r}+v_{\phi}B_{z}),
\\
&& \dot{v}_{\phi}=\frac{q}{m}(v_{z}B_{r}-v_{r}B_{z}),
\\
&& \dot{v}_{z}=\frac{q}{m}(E_{z}-v_{\phi}B_{r}).
\end{eqnarray}
\end{subequations}
Electron motion in the z-axis can be reduced into $\dot{v}_{z}=qE_{z}/m$ because the contribution of the electric force is more significant than the magnetic force during the acceleration.
By differentiating $\dot{v}_{r}$, $\dot{v}_{\phi}$ and rewriting the equation, we have
\begin{subequations}
\begin{eqnarray}
&& \ddot{v}_{r}=\frac{q}{m}[\dot{E}_{r}+\frac{q}{m}B_{z}(v_{z}B_{r}-v_{r}B_{z})],
\label{eq:fivea}
\\
&& \ddot{v}_{\phi}=\frac{q^{2}}{m^{2}}[(\textbf{\textrm{E}}\times\textbf{\textrm{B}})_{\phi}-v_{\phi}(B^{2}_{z}-B^{2}_{r})].
\label{eq:fiveb}
\end{eqnarray}
\end{subequations}
Note that the electric fields \textit{experienced} by a moving electron are not constant.
They are functions of positions which can be transformed to functions of time as shown in \crefformat{figure}{Fig.~#2#1{(c)}#3}\cref{fig:Figure3}.
The transverse electric component can further be expressed as a function of time: $E_{r}(z)=E_{r}(t)=E_{0}e^{i\omega t}$.
This relation can be obtained from simulations, where $\omega$ will be very close to the cyclotron frequency $\omega_{c}=qB_{z}/m$.  Eq.~(\ref{eq:fivea}) and (\ref{eq:fiveb}) can be further simplified by merging contributions of the transverse magnetic field into the cyclotron motion.
It reads
\begin{subequations}
\begin{eqnarray}
&& \ddot{v}_{r}=-\omega^{2}_{c}(v_{r}-\frac{i\omega}{\omega_{c}}\frac{E_{0}e^{i\omega t}}{B_{z}}),
\\
&& \ddot{v}_{\phi}=-\omega^{2}_{c}[v_{\phi}+\frac{(\textbf{\textrm{E}}\times\textbf{\textrm{B}})_{\phi}}{B^{2}_{z}}].
\end{eqnarray}
\end{subequations}
The transverse velocity of a single electron motion with the non-uniform electric and magnetic fields can be decomposed as 
\begin{equation}
\begin{aligned}
\textbf{v}_{t}&\approx \textbf{v}_{o}+\textbf{v}_{E}+\textbf{v}_{p} \\
&=\textbf{v}_{\perp}e^{i\omega_{c}t}-\frac{\textbf{\textrm{E}}\times\textbf{\textrm{B}}}{B^{2}_{z}}+\frac{\dot{\textbf{E}}_{r}}{\omega_{c}B_{z}\mid 1-\omega^{2}/\omega^{2}_{c}\mid},
\end{aligned}
\label{eq:seven}
\end{equation}
where $\textbf{v}_{o}$ is the cyclotron velocity in the non-uniform magnetic field, $\textbf{v}_{E}$ is the $\textbf{E}\times\textbf{B}$ drift velocity, and $\textbf{v}_{p}$ is the resonant polarization drift velocity.
Traditionally, in the design of MIGs, only the cyclotron motion and the $\textbf{E}\times\textbf{B}$ drift are discussed.
For the commonly used MIGs, the polarization drift is ignorable because the change rate of time-varying electric field \textit{experienced} by the gyrating electrons is much larger than the cyclotron frequency ($\omega\gg\omega_{c}$).
However, the polarization drift can play an important role in the non-adiabatic process when the frequency of the time-varying electric field approximates to the cyclotron frequency, i.e., $\omega\approx\omega_{c}$.
Figure~\ref{fig:Figure4} decomposes all the velocities as shown in Eq.~(\ref{eq:seven}), where (a) the corrected cyclotron drift, (b) the $\textbf{E}\times\textbf{B}$ drift, (c) the resonant polarization drift, and (d) the overall transverse velocity.
Both the $\textbf{E}\times\textbf{B}$ drift and resonant polarization drift would increase the velocity spread in \textit{phase} 1 of the non-adiabatic process but significantly reduce the velocity spread in \textit{phase} 2.

\begin{figure}
\includegraphics[scale=0.7]{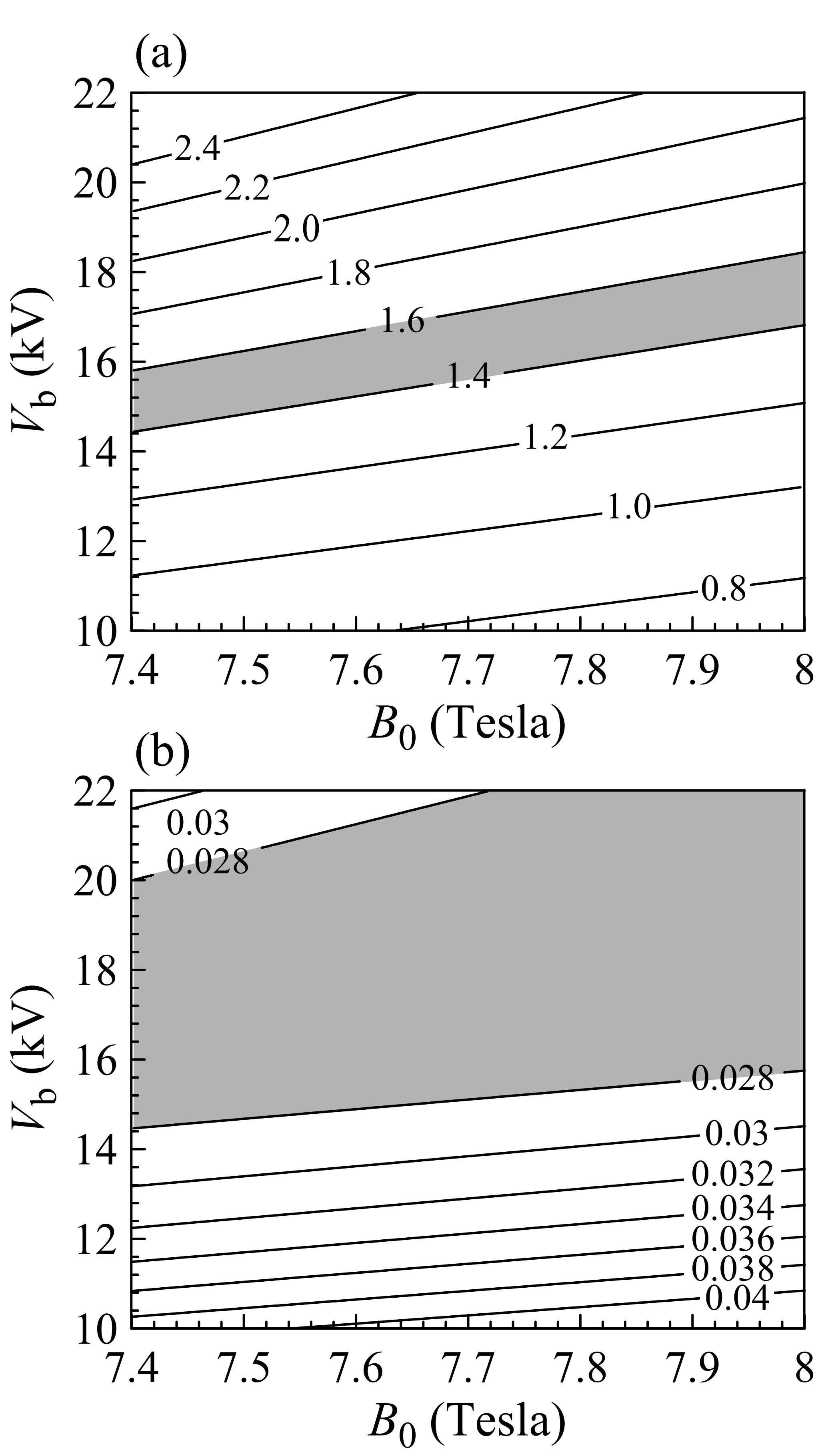}
\caption{\label{fig:Figure5}(a) The pitch factor and (b) the transverse velocity spread as functions of the beam voltage and the magnetic field.
The shadow areas mark the region of interest.}
\end{figure}

Figure~\ref{fig:Figure5} displays the simulated results based on the EGUN code verified by the CST particle studio.
All the operating parameters are listed in Tables~\ref{tab:table2}. The pitch factor and the transverse velocity spread of the electron beam as functions of the applied magnetic field and beam voltage are plotted in \crefformat{figure}{Fig.~#2#1{(a)}#3}\cref{fig:Figure5} and \crefformat{figure}{#2#1{(b)}#3}\cref{fig:Figure5}, respectively. The beam voltage and applied magnetic field have a simple and linear relation with the pitch factor as follows:
\begin{equation}
\alpha=1.5+0.135\times(V_{b}-16)-0.63\times(B_{0}-7.7),
\end{equation}
where $B_{0}$ is in the unit of Tesla and $V_{b}$ is in the unit of kV. 
The transverse velocity spread, on the other hand, has a complex dependence on both $B_{0}$ and $V_{b}$ with a relation:
\begin{equation}
\begin{aligned}
\Delta v_{\perp}/v_{\perp}(\%)=2.6&+0.0215\times(V_{b}-19)^{2} \\
&+0.3\times(B_{0}-8)^{2} \\
&-0.125\times(V_{b}-19)(B_{0}-8).
\end{aligned}
\end{equation}
The transverse velocity spread is small and good enough for the applications of frequency-tunable gyrotrons.
Adjusting either $V_{b}$ from 15 to 17 kV or $B_{0}$ from 7.4 to 8.0 T will maintain a high pitch factor of 1.5 and a minimal transverse velocity spread of 2.8$\%$.
The working ranges for $V_{b}$ and $B_{0}$ are broad and can be further extended depending on the requirements of gyrotron operation.
Comparing with the previous MIG\cite{Kumar2017}, the proposed electron gun provides a decent beam quality.
By using the similar interacting structure \cite{Chen2010} we expect to obtain a peak efficiency of 30$\%$ with a 3-dB tuning bandwidth of 8.5 GHz for a beam current of 0.5~A. 

\begin{table}
\caption{\label{tab:table2}Gun parameters.}
\begin{ruledtabular}
\begin{tabular}{lc}
Beam voltage, $V_{b}$ & $12-22$ kV\\
Magnetic field, $B_{0}$ & $7.4-8.0$ T\\
Beam current, $I_{b}$ & $0.2-1.0$ A\\
Cathode loading, $J_{emi}$ & $0.9-4.5$ A/cm$^{2}$\\
Cathode inclination, $\theta_{e}$ & 84$^{\circ}$\\
Peak electric field & 46 kV/cm\\
Emitter electric field & $23-29$ kV/cm\\
Emitter radius, $r_{c}$ & 3.5 mm\\
Emitter length, $l_{emi}$ & 1.0 mm\\
Beam radius, $r_{b}$ & 0.44 mm\\
Pitch factor, $\alpha$ & $1.0-2.4$\\
Transverse velocity spread, $\Delta v_{\perp}/v_{\perp}$ & $2.8-3.6\%$
\end{tabular}
\end{ruledtabular}
\end{table}

In conclusion, we proposed an unconventional cathode structure featuring a nearly planar emitter surface and a notch on the concave cathode, where the electrons have a much longer acceleration period, called the non-adiabatic process.
Sensitivity analysis shows that the proposed electrode structures provide a very stable operation condition with high machining tolerance.
The astonishing difference between theory and simulation is explained from the adiabatic and non-adiabatic viewpoints.
The $\textbf{E}\times\textbf{B}$ drift and the resonant polarization drift account for the low velocity spread under such a non-uniform and weak electric field.
This diode-type electron gun can produce an electron beam with the high pitch factor of 1.5 and the low transverse velocity spread of 2.8$\%$ over a very wide parameter space spanned by the magnetic field ($7.4-8.0$ T) and the beam voltage ($12-22$ kV).
The proposed design frees the limitation that the emitter should place at uniform and maximal electric fields and points out that the non-adiabatic process may be beneficial and plays a critical role in the electron gun design.
\\
\\
This work is supported by the Ministry of Science and Technology of the Republic of China (Taiwan) under contract No. 107-2112-M-007-015-MY3. The authors are grateful to Prof. W. Y. Chiang and CST Taiwan for technical support.

\end{document}